\documentclass[%
	times%
        ,5p%
        ,twocolumn%
	,sort&compress%
	]{elsarticle}
\journal{arXiv}

\usepackage{graphicx}

\usepackage{amssymb}
\usepackage{amsmath} 

\newcommand{\bvec}[1]{\boldsymbol{#1}} 
\newcommand{\KpEmpty}{K^{\emptyset}}
\newcommand{\Kp}[1]{K^{\{#1\}}}

\begin{document}

\begin{frontmatter} 

\title{Topological adiabatic dynamics in classical mass-spring chains with clamps}

\author{Atushi Tanaka}
\ead[url]{https://arxiv.org/a/tanaka\_a\_1.html}

\address{Department of Physics, Tokyo Metropolitan University,
  Hachioji, Tokyo 192-0397, Japan}

\begin{abstract}
  The path dependence of adiabatic evolution in classical harmonic
chains with clamps is examined. It is shown that cutting and joining a
chain may braid adiabatic normal mode frequencies. Accordingly,
different adiabatic paths with the same endpoints may transport a
normal mode to a different one, and an adiabatic cycle pumps action
variables, i.e., the adiabatic invariants of integrable classical
systems. Another adiabatic pump for artificial edge modes induced by
clamps is shown as an application. Extensions to completely integrable
systems and quantum systems are outlined.
\end{abstract}








\end{frontmatter} 

\section{Introduction}
\label{sec:Introduction}
The adiabatic theorems of classical and quantum mechanical systems tell us the presence of conserved quantities, i.e., invariants, during the time evolution where system parameters are slowly varied. In quantum theory, the populations of stationary states are adiabatic invariants.
In classical mechanics, the adiabatic invariants of integrable and ergodic systems are actions~\cite{GoldsteinAT} and phase-space volumes~\cite{Ott-PRL-42-1628,Jarzynski-PRL-71-839}, respectively.
The adiabaticity has been applied variously. To name a few, the adiabatic invariance of the classical action was a key to developing the quantum theory, and the quantum adiabatic theorem offers bases for studying molecular physics~\cite{Born-AnnPhysik-389-457,BornHuangAdiabatic} and quantum computation~\cite{Farhi-quant-ph-0001106,Kadowaki-PRE-58-5355,Suzuki-JPSJ-74-1649}.
Note that we here discern the adiabaticity in thermodynamics from the one in mechanics and focus on the latter.

An adiabatic process induces changes in physical quantities, and some are associated with the geometry of the adiabatic path in the parameter space.
Famous examples are the geometric phase factors of quantum states and the angle variable holonomy in classical integrable systems~\cite{Berry-PRSLA-392-45,Simon-PRL-51-2167,Aharonov-PRL-58-1593,Hannay-JPA-45-066501,Bohm-GPQS-2003,Chruscinski-GPCQM-2004}.
The geometric phase factor is crucial in understanding the quantum Hall effect~\cite{Thouless-PRL-49-405} and topological pumps of quantum~\cite{Thouless-PRB-27-6083} as well as classical~\cite{Kraus-PRL-109-106402,Rosa-PRL-123-034301,Cheng-PRL-125-224301} systems.

In quantum theory, the adiabatic invariants, i.e., the populations of stationary states, can also be changed by adiabatic processes. Namely, an adiabatic cycle can transport a stationary state to another stationary state~\cite{Cheon-PLA-248-285,Tanaka-PRL-98-160407,Yonezawa-PRA-87-062113}. The result of the adiabatic evolution of a stationary state is path-dependent, and the topology of the path determines the final state~\cite{Tanaka-PLA-379-1693,Tanaka-FAOTQP-531}. Recently, its experimental realization in a one-dimensional dipolar gas has been dubbed a quantum energy pump~\cite{Kao-Science-371-296}.

On the other hand, in classical mechanics, it has been unknown whether the adiabatic invariants can be path-dependent or, equivalently, the pumping of the classical adiabatic invariants is possible.
Although the classical energy pump studied by Lu, Jarzynski, and Ott~\cite{Lu-PRE-91-052913} appears to be such an example, their analysis proved that the pumping in their model is nonadiabatic.
The primary difficulty in finding a classical adiabatic invariant pump would be the absence of quantum-classical correspondence of the adiabatic theorem in generic systems, whose phase-space involves both regular and chaotic components~\cite{Ott-PRL-42-1628,Lochak-MultiphaseAveraging}.
Also, in classically integrable systems, an adiabatic path must avoid separatrix crossings to protect the adiabatic invariance of action variables~\cite{Neishtadt-SovPhysDokl-20-189,Neishtadt-JApplMathMech-39-594,Cary-PRA-34-4256}.

Our goal here is to show that the adiabatic evolution of classical adiabatic invariants is path-dependent in harmonic chains, which are completely integrable systems with multiple degrees of freedom.
Crucial ingredients are adiabatic ``cutting'' and ``joining'' operations to braid adiabatic normal mode frequencies.
We note that similar operations provide an example of path-dependent quantum adiabatic invariants~\cite{Cheon-ActaPolytechnica-53-410}.

This paper is organized as follows. In section~\ref{sec:model}, we examine mass-spring chains with clamps to prove the path dependence of the adiabatic invariants. An adiabatic cycle accordingly pumps the normal modes of a mass-spring chain (Section~\ref{sec:modepump}).
These points reflect the topological nature of adiabatic dynamics in the classical integrable systems (Section~\ref{sec:holonomyinaction}).
As an application, we introduce a topological pump for artificial edge modes induced by clamps (Section~\ref{sec:edgepump}).
In section~\ref{sec:summary}, we present a summary of the present work. We also briefly discuss the extension of our results to completely integrable classical systems and quantum systems. Appendices provide details of the adiabatic paths for our examples.

\section{Mass-spring chains with adiabatic pinning}
\label{sec:model}
\begin{figure}
  \centering
  \includegraphics[%
	width=8.66cm
  ]{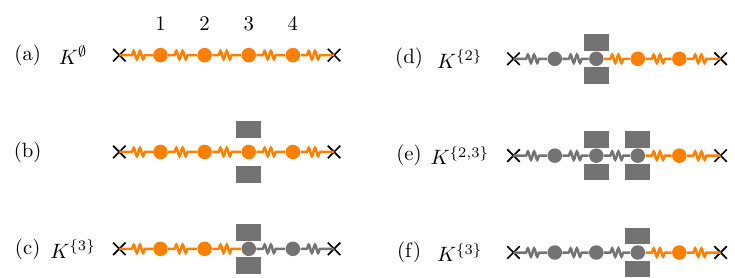}
  \caption{%
    The adiabatic time evolutions of a mass-spring system ($N=4$) along two paths $C_1$ and $C_2$, which share endpoints.
    (a) The mass-spring chain has four particles (indicated by circles) and satisfies the fixed boundary condition (indicated by two crosses). The system is initially in the slowest normal mode.
    (b) During $C_1$, the third particle is adiabatically clamped (indicated by two rectangles).
    (c) After completing $C_1$, the oscillation localizes at the left part of the system, and thus the right part (indicated by gray circles) ceases.
    During $C_2$, the second and third particles are clamped successively ((d) and (e)), and the clamp for the second particle is released finally. The resultant oscillation localizes at the right part of the system (f).
    The values of the adiabatic parameter $K$ (e.g., $\KpEmpty$ and $\Kp{3}$) are explained in the main text.
  }
  \label{fig:ms_open3}
\end{figure}
We examine classical mass-spring chains to explain the path-dependence of adiabatic evolution.
Assume that $N$ particles of identical mass $m$ are connected by harmonic springs of identical spring constant $k$. We impose the fixed boundary condition on the chain.
Each particle is imposed pinning potential adiabatically.
We assume that the system is described by the Hamiltonian
\begin{equation}
  \label{eq:defH}
  H = \frac{1}{2}\sum_{i=1}^N p_i^2 + \frac{1}{2}\sum_{i,j=1}^N K_{ij}x_i x_j
  ,
\end{equation}
where $x_i$ and $p_i$ are the displacement and momentum of the $i$-th particle ($1\le i\le N$) and $m=1$ is assumed. The $N\times{}N$ real symmetric matrix $K$ describes the harmonic springs and pinnings and is adiabatically varied.
According to the adiabatic theorem, once the system is prepared to be in a normal mode, specified by a normalized eigenvector $\bvec{\xi}_a$ of $K$, the system stays in a normal mode.
Let $\omega_a$ denote the normal mode frequency corresponding to $\bvec{\xi}_a$.

Our adiabatic paths visit the points where the particles are clamped to bring the braiding of adiabatic normal modes. The clamps are made of infinite pinning strength. For example, $i$-th particle is clamped when $K_{ii}$ is infinitely large. We denote $K=\Kp{i}$ if $i$-th particle is clamped and no pinning is imposed on other particles.
We note that a mathematical analysis of the parametric evolution along variable pining strength is shown in Ref.~\cite{ArnoldOscillation} to examine classical oscillations with variable rigidity.
Also, similar operations have been utilized to study the role of topology change in quantum systems~\cite{Balachandran-NPB-446-299,Shapere-1210.3545}.

We introduce the endpoints of adiabatic operations.
For simplicity, $N=4$ is assumed.
At the initial point of the paths, no pinning is imposed on the mass-spring chain, denoted as $K=\KpEmpty$ (Fig.~\ref{fig:ms_open3} (a)).
On the other hand, we assume $K=\Kp{3}$, i.e., infinitely strong pinning is imposed to clamp the third particle, at the endpoint.
%
%
The system at $\Kp{3}$ is accordingly divided into two, where the ``left'' and ``right'' parts are $N=2$ and $N=1$ mass-spring chains under the fixed boundary condition, respectively (Fig.~\ref{fig:ms_open3} (c) and (f)).

Both of two adiabatic paths $C_1$ and $C_2$ connect $\KpEmpty$ and $\Kp{3}$.
During the path $C_1$, the pinning is imposed only on the third particle (Fig.~\ref{fig:ms_open3} (b)) to adiabatically cut the whole system (Fig.~\ref{fig:ms_open3}~(c)).
On the other hand, $C_2$ involves ``cut'' and ``splice''.
Namely, $C_2$ connects $\KpEmpty$, $\Kp{2}$, $\Kp{2,3}$ and $\Kp{3}$ in order, where $i$-th and $j$-th particles are simultaneously clamped at $\Kp{i,j}$ ($1\le{}i,j\le N$) (Figs.~\ref{fig:ms_open3} (a), (d), (e) and (f)).
Examples of the parametrization of the paths are shown in~\ref{sec:path4}.

\begin{figure}[t]
  \centering
  \includegraphics[%
	width=6.50cm%
  ]{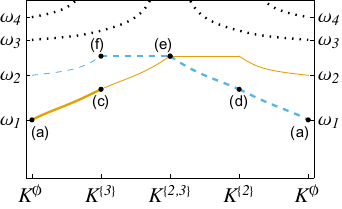}
  \caption{%
    Normal mode frequencies in $N=4$ mass-spring chain with clamps.
    The horizontal axis represents the closed path $C$, where the configurations of clamps are indicated.
    The adiabatic evolution of the ground normal mode $\omega_1$ at $\KpEmpty$ along $C_1$ (thick line) and $C_2$ (thick dashed line) emanate from the left and right ends, respectively, where the intermediate points shown in Fig.~\ref{fig:ms_open3} are depicted by circles.
    The evolutions of the ground mode from the left end (thick line) to the right end (thin line) arrive at the first excited mode $\omega_2$, where the continuity of the normal mode vector is kept.
    Hence the closed path $C$ induces the mode holonomy.
  }
  \label{fig:ms_close}
\end{figure}

We examine the adiabatic evolutions along $C_1$ and $C_2$.
We assume that the initial system at $K=\KpEmpty$ is in the ground normal mode, with frequency $\omega_1$ being the slowest.

The adiabatic evolution along $C_1$ delivers the ground mode of $\KpEmpty$ to the one of $\Kp{3}$.
This is due to the absence of spectral degeneracy in $C_1$ (see, Fig.~\ref{fig:ms_close}).
Since the system at $\Kp{3}$ is divided into two,
the final ground mode localizes at the larger, left part.
The rightmost particle accordingly ceases to oscillate although it is not clamped (Fig.~\ref{fig:ms_open3}~(c)).

On the other hand, the adiabatic time evolution along $C_2$ proceeds as follows.
The system at $\Kp{2}$ is in the ground mode and localized at the right side of the clamp (Fig.~\ref{fig:ms_open3} (d)) because of the symmetry between $\Kp{2}$ and $\Kp{3}$.
The adiabatic clamp on the third particle ``compresses'' the mode at $\Kp{2,3}$ to localize at the rightmost particle (Fig.~\ref{fig:ms_open3} (e)).

We must correctly handle the degeneracy point at $\Kp{2,3}$. The degenerate normal mode vectors localize at either end of the mass-spring chain. Note that these normal modes are not hybridized, as the complete clamps separate them. The system's state accordingly localizes at the right part around $\Kp{2,3}$ in $C_2$. Also, the state arrives at the first excited normal mode after $\Kp{2,3}$ and stays there. Hence the adiabatic evolution along $C_2$ transports the ground mode at $\KpEmpty$ to the first excited mode at $\Kp{3}$.

We conclude that the result of the adiabatic evolution of a normal mode from $\KpEmpty$ to $\Kp{3}$ depends on the path.
Furthermore, the adiabatic evolution is topological, as the possible candidates of the final normal modes are discrete.

\section{Normal mode pump}
\label{sec:modepump}
We utilize the path-dependent adiabatic time evolution to construct a normal mode pump.
Let $C$ denote the closed path made of $C_1$ and the inverse of $C_2$.
Namely, $C$ connects $\KpEmpty$, $\Kp{3}$, $\Kp{2,3}$, $\Kp{2}$ and $\KpEmpty$ in order. The adiabatic evolution along $C$ transports the ground mode at $\KpEmpty$ to the first excited mode at $\KpEmpty$, according to the analysis of $C_1$ and $C_2$ above (see, Fig.~\ref{fig:ms_close}).

\section{Multiple-valued adiabatic invariants}
\label{sec:holonomyinaction}
We examine the adiabatic invariants under the operation of the normal mode pump.
A normal mode of the mass-spring chain has a corresponding action variable.
The normal coordinate $Q_a$ and its canonical momentum $P_a$ are determined as $Q_a = \sum_i \xi_{ai}x_i$ and $P_a=\sum_a \xi_{ai}p_i$,
where $\xi_{ai}$ is the $i$-th component of the normal mode vector $\bvec{\xi}_a$.
The Hamiltonian~\eqref{eq:defH} is accordingly expressed as
$H=\frac{1}{2} \sum_a\left(P_a^2 + \omega_a^2 Q_a^2\right)$.
The adiabatic invariants of this system are the action
$J_a=\pi\left(P_a^2 + \omega_a^2 Q_a^2\right)/\omega_a$.

We explain the adiabatic evolution of the ground mode action $J_1$ during the cycle $C$.
Let $J_1^{(0)}$ denote the value of the action $J_1$ at the the initial point $\KpEmpty$. The value of $J_1$ is equal to $J_1^{(0)}$, according to the adiabatic theorem. Note that the adiabatic invariance is not suffered by the mode degeneracy at $\Kp{2,3}$, as is explained above.

Meanwhile, as a function of $x_i$ and $p_i$, $J_1$ varies along $C$ because the functional dependence of $J_1$ on $x_i$ and $p_i$ is determined through the parametric dependence of the normal mode vector $\bvec{\xi}_1$ on the adiabatic parameters.
As the normal mode pump along $C$ transports $\bvec{\xi}_1$ to $\bvec{\xi}_2$, the function $J_1$ is also transported to $J_2$.
Hence $J_2^{(1)}$, the value of $J_2$ at the end of $C$, is equal to $J_1^{(0)}$.
Now we see how the multiple-valuedness of the classical actions in the adiabatic parameter space is consistent with the adiabatic invariance of their values.

The example above suggests an amendment to the adiabatic theorem.
Here we assume that the adiabatic condition is satisfied for a cycle $C$ so that the action $J_a$ is an adiabatic invariant~\cite{GoldsteinAT}.
Let $J_a^{(0)}$ and $J_a^{(1)}$ denote the values of $J_a$ at the initial and final points of $C$, respectively.
The adiabatic cycle $C$ transports $J_a$ to $J_{s(C;a)}$,
where the index $s(C;a)$ may take the original value $a$, or another value, which depends on the initial index $a$ and the cycle $C$.
Hence $J_{s(C;a)}^{(1)}=J_a^{(0)}$ holds at the end of the cycle. In other words, we obtain
\begin{equation}
  \label{eq:permAction}
  J_a^{(1)}=\sum_b S_{ab}(C) J_b^{(0)}
  ,
\end{equation}
where $S_{ab}(C)=\delta_{a,s(C;b)}$ is $(a,b)$-th element of permutation matrix $S(C)$, which describes holonomy in action.

The action permutation $S(C)$ is determined by a topological nature of the adiabatic cycle $C$ because the permutation matrix $S(C)$ is discrete-valued.
As for the systems of small oscillation (Eq.~\eqref{eq:defH}), we may determine $S(C)$ from the adiabatic evolution of the normal mode vector $\bvec{\xi}_a$, which is a normalized eigenvector of the matrix $K$. The adiabatic evolution along $C$ transports $\bvec{\xi}_a$ to $\bvec{\xi}_{s(C;a)}$ up to a phase factor, which implies
$S_{ab}(C) = \left|\bvec{\xi}_{a}\cdot\bvec{\xi}_{s(C;b)}\right|$.
Hence the problem of determining $S(C)$ can be cast into the adiabatic response of a quantum system whose Hamiltonian is a symmetric matrix $K$. The adiabatic cycle $C$ transports a stationary state of the quantum system into itself or another stationary state, which is called exotic quantum holonomy\cite{Cheon-PLA-248-285,Tanaka-PRL-98-160407}.
It is known that an energy level crossing, which amounts to a crossing of normal mode frequency as for the harmonic chain, is a key to inducing exotic quantum holonomy in Hamiltonian systems~\cite{Cheon-PLA-374-144,Zhou-PRB-94-075443,Spurrier-PRA-97-043603}.

We add remarks on using infinitely strong clamps, which is the key to realizing path-dependent adiabatic evolution and mode pumping.
%
%
The mass-spring chain is divided into fragments at the limit of infinitely strong clamps. Because von Neumann and Wigner's non-crossing theorem is inapplicable~\cite{Cheon-ActaPolytechnica-53-410,Harshman-PRA-95-053616}, the normal mode spectrum may exhibit degeneracies.
Also, an infinitely strong clamp induces a divergent normal mode whose eigenvector localizes at the clamped particle.
Examples are $\omega_3$ and $\omega_4$ in FIG~\ref{fig:ms_close}.
The divergent modes need to be excluded from the sum in Eq.~\eqref{eq:permAction} since the adiabatic evolution that involves the divergent modes is not well-defined.

On the other hand, an imperfect clamp lifts level crossings, implying that the actions' destinations are qualitatively different from the one in the strict adiabatic limit. Still, the mode pumping occurs if the velocity of the parameter is suitably chosen, referred to as the diabatic evolution~\cite{Lichten-PR-131-229,Smith-PR-179-111}.

\section{Pumping artificial edge modes}
\label{sec:edgepump}
By reorganizing the cycle for the mode pump, we introduce a cycle $C_{\rm ee}$ for another mode pump that permutates localized normal modes.
Namely, the adiabatic cycle $C_{\rm ee}$ transports the oscillatory motion localized on a particle into another particle in a distant place.
Such a transport resembles the one in the Thouless pump~\cite{Thouless-PRB-27-6083}, which has mechanical and optical variants~\cite{Kraus-PRL-109-106402,Rosa-PRL-123-034301,Cheng-PRL-125-224301}.

The initial point of the localized mode pump $C_{\rm ee}$ is $\Kp{2,3}$, where the second and third particles are completely clamped, and the ground normal mode frequency doubly degenerates (Fig.~\ref{fig:ms_close}).
The corresponding normal mode vectors localize at either the leftmost or rightmost particle (e.g., Fig.~\ref{fig:ms_open3} (f)).
The cycle $C_{\rm ee}$ connects $\Kp{2,3}$, $\Kp{3}$, $\KpEmpty{}$, $\Kp{2}$ and $\Kp{2,3}$. The direction of the cycle is opposite to the cycle $C$ (Fig.~\ref{fig:ms_close}).
\begin{figure}[t]
  \centering
  \includegraphics[%
	width=6.50cm%
  ]{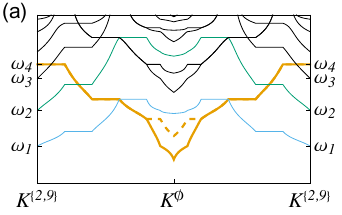}
  \includegraphics[%
	width=6.50cm%
  ]{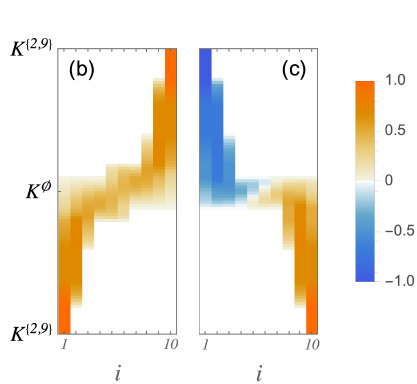}
  \caption{%
    An adiabatic topological pump in $N=10$ mass-spring chain with clamps along a closed path, which connects $\Kp{2,9}$, $\KpEmpty$, and $\Kp{2,9}$.
    (a) The parametric evolution of normal mode frequency. The pumping modes emanate from $\omega_4$ at the initial point and are doubly degenerated except around $\KpEmpty$ (thick line and thick dashed-line).
    (b) The density plot of the $i$-th component of a normalized pumping mode vector. The horizontal and vertical axes represent the particle number $i$, and the position at the cycle, respectively. The cycle transports the leftmost edge mode to the rightmost edge mode via the ground mode around $\KpEmpty$.
    (c) The density plot of another pumping mode vector, which is the first excited mode around $\KpEmpty$.
    For details, see~\ref{sec:path10}.
  }
  \label{fig:nmp10}
\end{figure}

We explain that the adiabatic cycle $C_{\rm ee}$ varies the initial leftmost edge mode into the rightmost one.
During the initial point $\Kp{2,3}$ to $\Kp{3}$,
the left part of the mass-spring chain is adiabatically expanded. Hence the system at $\Kp{3}$ is in the ground normal mode that is localized at the left room (see, Figs~\ref{fig:ms_open3}(c) and~\ref{fig:ms_close}(c)). After that, the system stays in the ground mode since there is no degeneracy except at the endpoint $\Kp{2,3}$. In particular, during the latter half of the path, i.e., $\KpEmpty{}$ to $\Kp{2,3}$, the oscillation is compressed to the rightmost particle. Hence the final state is the rightmost edge mode.

Similarly, $C_{\rm ee}$ pumps the rightmost edge mode to the leftmost one. Note that these two edge-edge pumping processes are not symmetric. The system's state stays at the first excited normal mode, which is different from the one during the opposite process, except at the endpoints of $C_{\rm ee}$.

Although $C_{\rm ee}$ does not pump the frequency of the normal mode, it induces the ``rotation'' within the normal mode eigenspace. This resembles the Wilczek-Zee holonomy~\cite{Wilczek-PRL-52-2111}.
This rotation, in reality, is due to an eigenspace permutation that occurs at the non-degenerate region in the adiabatic path, so the holonomy is somewhat exotic~\cite{Cheon-PLA-248-285,Tanaka-PRL-98-160407}\footnote{%
  A similar rotation has already been examined in a study of the adiabatic quantum control of atomic and molecular systems~\protect\cite{Yatsenko-PRA-65-043407}.
}. 
The action permutation induced by the edge-edge pump $C_{\rm ee}$ can be described by Eq.~\eqref{eq:permAction}.

Extending the present scheme to larger mass-spring chains is possible as long as the adiabatic time evolution is ensured, e.g., the density of mode frequencies is sparse enough.
An adiabatic topological pump in $N=10$ mass-spring chain
is shown in Fig.~\ref{fig:nmp10},
whose details
%
are explained in Appendix~B.

Our edge-edge pump differs from the edge-edge pumps based on the Thouless pump in mechanical systems~\cite{Rosa-PRL-123-034301}.
The design of the Thouless edge-edge pumps is based on the band structure of the underlying periodic system.
The topological nature of the system ensures the presence of boundary states, which are the targets for the pump~\cite{Kraus-PRL-109-106402,Rosa-PRL-123-034301,Cheng-PRL-125-224301}.
Meanwhile, in the present case, we do not need to consult any underlying periodic system, and the infinitely strong clamps enforce the presence of artificial edge modes.

\section{Summary and discussion}
\label{sec:summary}
We have examined how the normal modes of the mass-spring chain with clamps are adiabatically evolved. It is shown that the destination may depend on the path even when the initial normal mode and the endpoints of paths are given. In particular, when the path is closed, the adiabatic actions corresponding to normal modes are permutated (Eq.~\eqref{eq:permAction}).
Because the present example of holonomy in action is the exchange of actions between degrees of freedom, it only occurs in multiple degrees of freedom systems.
In other words, the mechanism behind the action pumps in single-degree of freedom classical systems must differ from the one presented here~\cite{tcip}.

Since the adiabatic mode evolution can be understood in terms of the eigenvalue problem of parameter-dependence of $K$ matrix (Eq.~\eqref{eq:defH}), the formulations for eigenspace permutations in quantum systems~\cite{Viennot-JPA-42-395302,MehriDehnavi-JMP-49-082105,Tanaka-PLA-379-1693,Tanaka-FAOTQP-531,Pap-Sigma-18-033} are applicable if the degeneracies in normal mode frequencies are appropriately treated.
Also, since an adiabatic open path may permutate eigenfunctions of quantum systems without spectral degeneracies~\cite{Lauber-PRL-72-1004,Pistolesi-PRL-85-1585,Manini-PRL-85-3067}, it is worth finding an open adiabatic path for topological pumping, which could circumvent the subtleties induced by degeneracies.

It is straightforward to extend the present work, e.g., the action permutation induced by adiabatic cycles (Eq.~\eqref{eq:permAction}), to completely integrable systems.
Still, finding examples in nonlinear systems seems challenging because the system needs to be completely integrable throughout an adiabatic path, and the system's state needs to avoid separatrix crossings~\cite{Neishtadt-SovPhysDokl-20-189,Neishtadt-JApplMathMech-39-594,Cary-PRA-34-4256}.

On the other hand, extending the present result to quantum mass-spring chains is straightforward. With the help of the semiclassical quantization for the classical actions~\cite{SemiclassicalQuantization} in Eq.~\eqref{eq:permAction}, we obtain the permutation of quantum number induced by an adiabatic cycle $C$:
\begin{equation}
  \label{eq:permNumber}
  n_a^{(1)}=\sum_b S_{ab}(C) n_b^{(0)}
  ,
\end{equation}
where $n_a^{(0)}$ and $n_a^{(1)}$ are the values of the quantum number for $a$-th mode at $C$'s initial and final points, respectively.
In contrast with the classical case, such a quantum number holonomy would be stable against small perturbations that may break the complete integrability in the underlying classical system. Also, the quantization procedure provides us with a quantum topological pump based on the holonomy in action.

\section*{Acknowledgments}
The author would like to thank Hiroaki Imada for collaboration at an early stage of this work, Professor Michael Wilkinson for bringing attention to Ref.~\cite{Thouless-PRB-27-6083}, and Kenji Komorida for discussion.
This work has been supported by JSPS KAKENHI Grant Number 20K03791.

\appendix
\section{Adiabatic dependence of pinning potentials for $N=4$ mass-spring chain}
\label{sec:path4}
We explain how we parameterize the adiabatic paths for $N=4$ mass-spring chain with clamps.
Let $s$ be the adiabatic parameter, which determines the value of
the $K$ matrix in Eq.~\eqref{eq:defH}. In the present case, $K$ is $4\times4$
and consists of $s$-independent term $\KpEmpty$ and $s$-dependent diagonal matrix $\left\{K_{ii}^{{\rm p}}(s)\right\}_{i=1}^{4}$. The former describes the free mass-spring chain, and the latter describes the pinning potential.
Here, the pinning is only imposed on the second and third particles, i.e., $K_{11}^{{\rm p}}(s)=K_{44}^{{\rm p}}(s)=0$.

First, we explain the open path $C_1$ that connects $\KpEmpty$ and $\Kp{3}$ (see, Figs.~\ref{fig:ms_open3} and~\ref{fig:ms_close}),
where the second particle is not clamped, i.e., $K_{22}^{{\rm p}}(s)=0$
and only the third particle is adiabatically clamped.
We impose the range of the adiabatic parameter $s$ as $[0,1/4]$.
%
An example of the clamp on the third particle is
\begin{align}
  K_{33}^{{\rm p}}(s)
  =\tan\left(2\pi s\right)
\end{align}
to ensure $K_{33}^{{\rm p}}(1/4)=\infty$ at the end of $C_1$.
Other parametrizations work as long as the adiabatic condition holds.

Second, we explain the cycle $C$, where $0\le s\le 1$ is assumed. We assume that $C_1$ and $C$ are the same during the initial stage $0\le s\le 1/4$.
The interval $1/4\le s\le 1/2$ in $C$ connects $\Kp{3}$ and $\Kp{2,3}$.
Here, the second particle is adiabatically clamped while the third particle is kept clamped completely, i.e.,
$K_{33}^{{\rm p}}(s)=\infty$.
%
The latter half of $C$ is for unclamp operations.
To connect $\Kp{2,3}$ and $\Kp{2}$ during $1/2\le s\le 3/4$, we impose
$K_{22}^{{\rm p}}(s)=\infty$ and
unclamp the third particle.
At last, $\Kp{2}$ and $\KpEmpty$ is connected during $3/4\le s\le 1$ to
unclamp the second particle.
An example of the precise parametrization for $C$ is
\begin{align}
  K_{22}^{{\rm p}}(s)
  =
  \begin{cases}
    0
    &\text{for}\quad 0\le s\le \frac{1}{4}\\
    \tan\left[2\pi \left(s-\frac{1}{4}\right)\right]
    &\text{for}\quad \frac{1}{4}\le s\le \frac{1}{2}\\
    \infty
    &\text{for}\quad \frac{1}{2}\le s\le \frac{3}{4}\\
    \tan\left[2\pi (1 - s)\right]
    &\text{for}\quad \frac{3}{4}\le s\le 1\\
  \end{cases}
\end{align}
and
\begin{align}
  K_{33}^{{\rm p}}(s)
  =
  \begin{cases}
    \tan\left(2\pi s\right)
    &\text{for}\quad 0\le s\le \frac{1}{4}\\
    \infty
    &\text{for}\quad \frac{1}{4}\le s\le \frac{1}{2}\\
    \tan\left[2\pi \left(\frac{3}{4} - s\right)\right]
    &\text{for}\quad \frac{1}{2}\le s\le \frac{3}{4}\\
    0
    &\text{for}\quad \frac{3}{4}\le s\le 1\\
  \end{cases}
  .
\end{align}

\section{Details of $N=10$ edge-edge pump}
\label{sec:path10}
We explain the adiabatic cycle for the edge-edge pump shown in Fig.~\ref{fig:nmp10}. The initial point of the cycle is $\Kp{2,9}$, where the normal mode $\omega_4$ is doubly degenerated, and the corresponding normal mode vectors are localized at the leftmost ($i=1$) and rightmost ($i=10$) particles.

These edge modes are adiabatically expanded during the former half of the cycle, where the initial point and $\KpEmpty$ are connected. The initial leftmost and rightmost edge modes are transported to the ground and the first excited modes at $\KpEmpty$ (Fig.~\ref{fig:nmp10} (b) and (c)), respectively, because the reflection symmetry is broken. The latter half of the cycle adiabatically compresses the ground and first excited modes at $\KpEmpty$ to produce the edge modes at the final point $\Kp{2,9}$.

In the following, our precise specification of the cycle is shown. There are four stages.

The first part of the cycle connects the initial point $\Kp{2,9}$, $\Kp{2,3,8,9}$, $\Kp{3,8}$, $\Kp{3,4,7,8}$ and $\Kp{4,7}$, where $\Kp{i,j,k,l}$ represents the case that $i$, $j$, $k$ and $l$-th particles are simultaneously clamped completely.
During this part, the edge modes are adiabatically expanded while the normal mode frequency is doubly degenerated. So the final states are localized at the left and right parts of the mass-spring chain.

Such ``broad edge modes'' are expanded further during the second part of the cycle that connects $\Kp{4,7}$, $\Kp{7}$ and $\KpEmpty$. The reflection symmetry is broken during this part to split the degeneracy in the normal mode frequency (see the bold and bold-dashed lines in Fig.~\ref{fig:nmp10}). The final states are the ground and the first excited modes.

The third part connects $\KpEmpty$, $\Kp{4}$ and $\Kp{4,7}$. This part can be an ``inverse'' of the second one, except that the way to break the reflection symmetry is slightly different. Hence the ground mode and the first excited modes become localized in the right and left rooms at $\Kp{4,7}$, where the frequency is doubly degenerated again.

The final part connects $\Kp{4,7}$, $\Kp{3,4,7,8}$, $\Kp{3,8}$, $\Kp{2,3,8,9}$ and $\Kp{2,9}$. This part is precisely the inverse of the first part to compress the broad edge modes.

The parametric dependence of the adiabatic clamp and unclamp operations is described by the tangent function, for example, as explained in~\ref{sec:path4}.

We explain the phases of normal mode vectors shown in Fig.~\ref{fig:nmp10}. We impose the parallel transport condition on these vectors. The components of the normal mode vector corresponding to the pumping from left to right are non-negative through the pumping cycle (Fig.~\ref{fig:nmp10}(b)). While the component of the normal mode vector corresponding to the pumping from right to left (Fig.~\ref{fig:nmp10}(c)) acquires the phase factor $-1$. These are the Manini-Pistolesi off-diagonal geometric phase factors, which can be understood from the normal mode vectors being eigenvectors of the $K$ matrix~\cite{Manini-PRL-85-3067,Tanaka-JPA-45-335305}.

\bibliographystyle{elsarticle-num}  


\end{document}